\documentclass[
unsortedaddress,
twocolumn,
amsmath,
amssymb,
aps,
prl,
]{revtex4-1}
\usepackage{braket}
\usepackage{graphicx}
\usepackage{lipsum}
\usepackage{dcolumn}
\usepackage{bm,color}
\usepackage[normalem]{ulem}

\usepackage{amsfonts}

\usepackage{epigraph}

\usepackage{dsfont}
\usepackage{etoolbox}
\makeatletter
\newlength\epitextskip
\pretocmd{\@epitext}{\em}{}{}
\apptocmd{\@epitext}{\em}{}{}
\patchcmd{\epigraph}{\@epitext{#1}\\}{\@epitext{#1}\\[\epitextskip]}{}{}
\makeatother

\begin{document}

\title{The non-Hermitian control of qubit dynamics}

\title{Coherent Qubit Cooling through Rabi Instantons}

\title{Real-Time Instantons in Complex-Driven Qubits}

\author{Samuel Alperin}

\affiliation{\vspace{1.25mm} \mbox{Los Alamos National Laboratory, Los Alamos, New Mexico 87545, USA}}

\begin{abstract}

We consider the dynamics of the quantum Rabi model driven parametrically by a periodic modulation of a complex coupling. We show both analytically and numerically that instead of Rabi oscillations, this nonunitary coherent driving leads to a unidirectional instanton solution which mediates the rapid and deterministic one-way tunneling of any initial coherent state to the ground state, making the ground state a strong  attractor in the quantum dynamics of the qubit. The timescale of this tunneling is shown to be inversely proportional to the effective resonant coupling, allowing for exceptionally fast, deterministic, and high-fidelity qubit reset through a purely coherent, PT-symmetric drive--without coupling to external dissipative baths, lossy resonators, or employing measurement-based feedback. Finally, we show how the drive can be engineered to place the strong attractor at any arbitrary point on the Bloch sphere.




\end{abstract}

\maketitle

One of the fundamental elements required for any universal quantum computing system is the initialization, or reset, of qubits \cite{Divincenzo2000}. Abstractly, the qubit reset is an operation which takes an arbitrary state as input, and gives a predefined state as the output -- such initializations are required prior to passing a qubit through any quantum logic circuits. 
However, beyond the initialization of qubits at the beginning of a quantum algorithm, it has been becoming increasingly clear that many reset operations are required for the implementation of error correcting schemes, which require the repeated readout and reset of auxiliary/syndrome qubits \cite{Magnard2018, Arute2021, Magnard2018}.
A number of early works on quantum information processing have relied on the conceptually simplest method of qubit reset, which passively exploits the natural tendency of a qubit to eventually relax into the ground state. However, given the increasing lifetimes of current qubit technologies as well as the need for reset operations that are fast and reliable enough not to preclude the use of error correcting schemes (which use many reset operations)  necessary for useful computations at scale, the qubit reset now represents the main limitation to quantum computing clock speed in many physical platforms \cite{Basilewitsch2021}.

As a result of the importance of qubit reset protocols, a number of active methods have been studied over the last decade. One approach is to read out the qubit state and then apply a tailored gate operation to rotate the qubit into the predetermined initial state \cite{Johnson2012, Riste2012}. However, by relying on active feedback, the speed and fidelity of readout-based approaches are limited by those of both the readout and subsequent gate operation. Other approaches, which do not require feedback, are generally based on tuning the coupling of the qubit to a lossy resonator to drive the qubit into the ground state faster than it would in isolation \cite{Magnard2018, Murch2012}. Such approaches typically require exploiting auxiliary states, which in turn requires going beyond the abstract treatment of a simple two-level quantum system, and which requires choosing between speed and fidelity \cite{Magnard2018, Ficheux2022, Murch2012}. Further, such approaches require coupling to the environment through incoherent processes, which allows for decoherence.

Meanwhile, the last two decades have witnessed an extraordinary rate of development in both the theory and physical realization of non-Hermitian quantum systems. In the time since it was realized that there exist certain non-Hermitian Hamiltonians with physically observable spectra (being strictly real and bounded below)\cite{Bender1998}, non-Hermitian quantum theory has led to the discovery of striking phenomena in both classical and quantum systems, such as unidirectional invisibility \cite{Lin2011} and topological lasing \cite{Bandres2018}. While non-Hermitian physics is not yet fully understood, and the full range of its associated phenomena is quite varied, a common theme appears to be the emergence of unidirectional transport. Such is the case of the original model of non-Hermitian physics, the Su-Schieffer-Heeger (SSH) model, a tight-binding model in which explicitly asymmetric couplings lead to asymmetric transport \cite{Su1979}, and as a result, the non-Hermitian skin effect \cite{Okuma2020}. However, recently it has been suggested that non-Hermitian systems might allow for another manifestation of unidirectional transport - that of mode occupation. For example, recent work in classical optics showed through detailed numerical experiments that unidirectional coupling could be achieved in multimode optical fibers using a periodic complex potential \cite{Akhter2023}. Thus, instead of the usual dynamics in such a fiber, in which an initially pure fundamental mode is unstable and deteriorates into random speckle, any input mode leads to a pure fundemental mode as the output.



In this Letter, we consider the quantum dynamics of a qubit under the complex parametric modulation of the cavity-dipole interaction. Specifically, we show through both analytical derivation and direct numerical simulations that under a certain kind of PT-symmetric parametric drive, the ground state of the qubit becomes a strong attractor in the quantum dynamics. We show that at the root of this phenomenon is the emergence of a unidirectional instanton solution which mediates the spontaneous tunneling from arbitrary initial states to the ground state. Thus under the non-Hermitian driving, a qubit in an unknown, arbitrary initial state can be sent to the ground state deterministically, and as we show, with fidelity that increases dramatically with time. 
This represents a novel and explicitly non-Hermitian form of qubit reset, which does not rely on feedback, auxiliary states, or any other structure beyond that of the abstract two-level system. These dynamics are in agnostic to the physical framework, and can readily be realized in neutral atom qubit systems, photonic qubit systems, and superconducting circuit based systems. For example, in the latter case, one can use standard transmon and resonator components, with the complex, time-dependent coupling $g(t)=g_{0}e^{-i\omega_{g}t}$ corresponding physically to a phase-controlled parametric modulation of the qubit--resonator impedance, achievable with presently available hardware such as flux-tunable couplers or Josephson mixers that implement active, gain--loss--balanced interactions at microwave frequencies. 
The required modulation rates (a few~GHz) and coupling strengths (tens of~MHz) lie well within the demonstrated range of state-of-the-art superconducting devices, indicating that the proposed PT-symmetric driving scheme is experimentally accessible using current technology.

We begin with the quantum Rabi model, a canonical model in quantum optics and the central object of study in cavity and circuit quantum electrodynamics,  which describes the dynamics of a two-level system coupled to a cavity mode \cite{Casanova2010, Bishop2010, Ermann2020}

\begin{equation}
    H_{R} = \omega_0 \hat{a}^\dagger \hat{a}+\omega_a\hat\sigma_{z}+g(\hat a^\dagger +\hat a)(\hat \sigma_{-} +\hat \sigma_+).
    \label{qrm}
\end{equation}
 While exact solutions to the quantum Rabi model were shown fairly recently to exist in terms of special functions \cite{Braak2011}, these solutions remain difficult to work with. Despite the discovery of those solutions, it has remained common to follow the method of Jaynes and Cummings \cite{Jaynes1963, GerryKnight2005}, which was once necessary for tractability. Their approximation is taken by writing the interaction component of the Hamiltonian in the Heisenberg picture as

\begin{equation} \label{jc0}
\begin{split}
H_{int} &= g(\hat a^\dagger e^{i\omega_0t} +\hat ae^{-i\omega_0t})(\hat \sigma_{+}e^{i\omega_qt} +\hat \sigma_-e^{-i\omega_qt})  \\
 & =g\Large{(}(\hat a \hat\sigma_+ +\hat{a}^\dagger\hat\sigma_-)e^{i(\omega_0-\omega_q)t}\\&\hspace{10pt}+(\hat a \hat\sigma_- +\hat{a}^\dagger\hat\sigma_+)e^{i(\omega_0+\omega_q)t}\Large{)},
\end{split}
\end{equation}
so that it becomes clear that for near-resonant conditions ($\omega_0\approx\omega_q$), $H_{int}$ can be separated into two timesscales. The slow dynamics are governed by the quantum excitation preserving operators $\{\hat a \hat\sigma_+ ,\hat{a}^\dagger\hat\sigma_-\}$, which both simultaneously annihilate (create) a cavity Boson and create (annihilate) a spin quanta. At very fast timescales, the dynamics are dominated by pseudoparticle processes, with the operators $\{\hat a \hat\sigma_- ,\hat{a}^\dagger\hat\sigma_+\}$ simultaneously annihilating (creating) both a cavity Boson and a spin excitation. As the latter processes do not contribute significantly to the long timescale dynamics of the system in the near-resonant, weak-coupling regime which is already natural to many experimental platforms, those terms, which are known as the counterrotating terms, can be neglected.  Returning to the Schrodinger picture, we are left with the celebrated Jaynes-Cummings (JC) model:

\begin{equation}
    H_{JC} = \omega_0 \hat{a}^\dagger \hat{a}+\omega_a\hat\sigma_{z}+g(\hat a^\dagger\hat \sigma_- +\hat a\hat \sigma_+)    \label{jcm}.
\end{equation}
This model, which is exactly integrable in terms of elementary functions, captures an incredible amount of physics, showing fundamental quantum cavity QED phenomena such as Rabi-oscillations. While it is difficult to overstate the effect that the approximation of Jaynes and Cummings has had on the field of quantum optics, recent advances in the development of cavity and circuit-QED systems which are able to delve increasingly into the deep-strong-interacting regime have required going beyond the JC model, in turn inspiring greater scrutiny of the physical effects of the counterrotating operators of the quantum Rabi model \cite{FornDiaz2019}.

The original conception of the quantum Rabi model was based on the physical picture of a two-level atom in an optical cavity. However, the same Hamiltonian can be realized with superconducting circuits, which couple microwave resonator photons to artificial atoms (superconducting qubits) \cite{Blais2021}. Perhaps the simplest version of this platform is constructed by capacitively coupling a transmon-like qubit (composed of a Josephson junction and a capacitor) to a linear resonator (a simple LC-circuit) \cite{Rasmussen2021}. In this simple system, the dynamics follow a quantum Rabi Hamiltonian with interaction $g\propto\sqrt{Z}$, where $Z$ is the impedance between the qubit and resonator. Using presently available superconducting circuit hardware, we note that this impedance can be modulated at microwave frequencies; the effects resulting from the parametric driving of the cavity-dipole interaction have thus been studied in the context of superconducting circuits. However, this impedance can also be made to be complex-valued, again using available superconducting circuitry which allow for quantities such as negative resistance/capacitance. Taking both time-dependence and complexity into account, one can consider a quantum Rabi interaction Hamiltonian of the form

\begin{equation} \label{jc0}
\begin{split}
H_{int} &= g(t)(\hat a^\dagger  +\hat a)(\hat \sigma_{+} +\hat \sigma_-),
\end{split}
\end{equation}
where $g(t):\mathbb{R}\rightarrow\mathbb{C}$. 
Considering the case of PT-symmetric complex parametric driving of the form $g(t)=g_0[\cos(\omega_g t)+i\sin(\omega_g t)]=g_0 e^{-i\omega_g t}$ -- exactly the functional form of the potential in the fully classical, PT-symmetric self-cleaning fiber discussed prior \footnote{Here PT (or \textit{Parity-Time}) symmetry refers a complex function being even in the real axis and odd in the imaginary axis.}. Rotating into the Heisenberg picture and writing $g(t)$ in exponential form, the interaction Hamiltonian takes the form

\begin{equation} \label{rab2}
\begin{split}
H_{int} &= \Large{(}(\hat a \hat\sigma_+ +\hat{a}^\dagger\hat\sigma_-)e^{i(\omega_0-\omega_q-\omega_g)t}\\&\hspace{10pt}+(\hat a \hat\sigma_- +\hat{a}^\dagger\hat\sigma_+)e^{i(\omega_0+\omega_q-\omega_g)t}\Large{)},
\end{split}
\end{equation}
Taking $\omega_g\approx \omega_0+\omega_q$ (sum-frequency parametric resonance), we find that there is again a natural separation of the Hamiltonian into two timescales. However, unlike in the case of the standard Jaynes-Cummings analysis, here it is the counter-rotating terms which dominate the slow-timescales, and the quanta-preserving co-rotating terms which can be time-averaged out. Under such time-averaging, we are left with a total Hamiltonian of the form

\begin{equation}
    H_{} = \omega_0\hat a\hat a^{\dagger}+\omega_a\hat\sigma_{z}+g_0(\hat a\hat \sigma_- +\hat a^{\dagger} \sigma_+)    \label{anti-jcm}.
\end{equation}
Defining the Casimir operator $\hat{C}=\hat a\hat a^{\dagger}+\frac{1}{2}(1-\hat\sigma_z)$, we can transform $\hat{b}=\hat a\hat\sigma_- /\sqrt{\hat C}$ and $\hat{b}^\dagger=\hat a^\dagger\hat\sigma_+/\sqrt{\hat C}$. After rescaling, this yields the Hamiltonian
\begin{equation}
    H_{} = \tilde\omega_0\hat{C}+\omega_a\hat\sigma_{z}+\tilde g_0(\hat b+\hat b ^\dagger)    \label{anti-jcm2}, 
\end{equation}
the terms of which now form a closed Lie algebra with nonzero commutators

\begin{equation}\label{comms}
\begin{split}
&[\hat\sigma_z,\hat b]= -2\hat b\\
&[\hat\sigma_z,\hat b^\dagger]= 2\hat b^\dagger\\
&[\hat b,\hat b^\dagger]=\Big(\frac{1}{\hat C}-1\Big)\hat \sigma_z.\ 
\end{split}
\end{equation}
Notably, while the Jaynes-Cummings model obeys an SU(2) algebra, here we find that the Rabi dynamics dominated by the counter-rotating terms are described by the SU(1,1) algebra, a closed yet noncompact algebra which is intimately tied to the nature of open dynamics.
By the theorems of Wei and Norman, the finite-dimensional closure of this algebra implies the existence of a solution to the quantum dynamics \cite{WeiNorman1963, WeiNorman1964} which takes the form of the time-evolution operator ansatz

\begin{equation}\label{ansz}
\begin{split}
   \hat{U}(t)&=\mathcal{T}\mathrm{exp}\left[\frac{-i}{\hbar}\int^t_0 \hat{H}(t)dt^{'}\right]\\&=e^{-if_0(t)\hat M}e^{-if_1(t)\hat{\sigma}_z}e^{-if_2(t)\hat{b}}e^{-if_3(t)\hat{b}^{\dagger}}.
\end{split}
\end{equation}
By noting that $-iH=\partial_t\hat U(t)\cdot \hat{U}^{-1}(t)$ and inserting both the Hamiltonian in \ref{anti-jcm} and the time-evolution anzatz in Eq. \ref{ansz}, the time-dependent functions $f_\alpha(t)$ can be constrained by the following set of ordinary differential equations
\begin{equation} \label{deq}
\begin{split}
\partial_tf_0 & =\omega_0
\\\partial_tf_1 &=i\alpha \cdot(1+f_1(t)^2)
\\\partial_tf_2 & = i\alpha\cdot(1+f_1(t)f_2(t))
\\ \partial_tf_3& =-i\alpha f_2
\end{split}
\end{equation}
which can each be solved exactly as
\begin{equation} \label{sols}
\begin{split}
f_0 & =\omega_0 t
\\f_1 &=i\tanh(\alpha t)
\\f_2 & = i\tanh(\alpha t)
\\ f_3& = \frac{1}{2}\log(\cosh(\alpha t)^2)
\end{split}
\end{equation}
Here we are primarily interested in the dynamics of the qubit spin, which is where quantum information is held. The time dependent form of the z-component of the qubit spin is defined as $\hat\sigma_z(t)=\hat{U}(t)\hat{\sigma}_z\hat{U}^{-1}(t)$. Inserting the solved time-evolution operator and taking a Baker-Campbell-Hausdorff expansion to second order yields the expectation value
\begin{equation}
\begin{split}\label{inst}
    \langle \hat\sigma_z \rangle(t)&=\hat\sigma_z(t_0)(1+2f_1(t)f_2(t))\\
    &=\hat\sigma_z(t_0)(1-2\tanh(\alpha t)).
\end{split}
\end{equation}

\begin{figure}[t]
\centering
\includegraphics[width=0.33\textwidth]{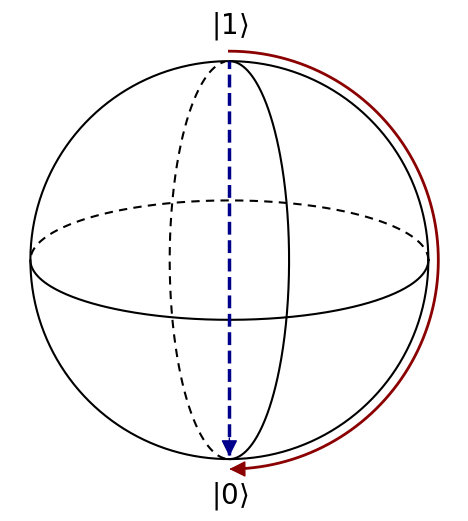}
\caption{
Conceptual sketch of the qubit trajectories on the Bloch sphere.
The red curved arrow shows the \emph{unitary Rabi path}, a surface rotation that connects the excited $\ket{1}$ and ground $\ket{0}$ states through coherent oscillation.
The blue dashed arrow shows the \emph{Bloch-instanton path}, the straight-line trajectory generated by the PT-symmetric, non-Hermitian drive.
In this case the azimuthal degree of freedom is suppressed, and the qubit follows an interior gradient-flow path directly through the Bloch ball—from the north to the south pole—realizing the real-time analog of a field-theoretic instanton.
}\label{cool}
\end{figure}
Remarkably, the evolution of the spin expectation value in Eq.~(12)
coincides exactly with the classical trajectory of a $\phi^4$ kink instanton.
Defining the auxiliary field $x(t)=(1-\langle\hat\sigma_z(t)\rangle)/2$,
one finds that it obeys
\begin{equation}
\dot x=\alpha(1-x^2)=-\frac{\partial}{\partial x}\,U(x),\qquad
U(x)=-\alpha\!\left(x-\frac{x^3}{3}\right),
\label{eq:instantonEOM}
\end{equation}
which is the first-order instanton equation in a double-well potential with fixed points
$x=\pm1$.  Under Wick rotation $t\!\to\! i\tau$ this becomes the Euclidean
instanton equation
\begin{equation}
\frac{dx}{d\tau}=\pm\sqrt{2V(x)},\qquad
V(x)=\frac{\alpha^2}{2}(1-x^2)^2,
\end{equation}
with finite Euclidean action $S_E=4\alpha/3$.
Hence $x(\tau)=\tanh(\alpha\tau)$ is the standard $\phi^4$ instanton/kink,
and $\langle\hat\sigma_z(t)\rangle=1-2x(t)=1-2\tanh(\alpha t)$
is its affine image in the qubit observable.

This affine transformation is geometrically natural:
$x=(1-z)/2$ is simply the population coordinate,
$p_e(t)=x(t)$ and $p_g(t)=1-x(t)$,
linked to the Bloch $z$-coordinate by $z=p_g-p_e$.
In the Bloch-sphere representation, 
\begin{equation}
z=\cos\theta,\qquad
x=\frac{1-z}{2}=\sin^2\!\frac{\theta}{2},
\end{equation}
so $x$ is the polar-cap area coordinate while $z$ measures the polar projection.
The $\phi^4$ instanton in $x$ therefore projects affinely to a monotonic
trajectory in $z$, representing motion along the $z$ axis of the Bloch ball.
Because the PT-symmetric, non-Hermitian drive suppresses the azimuthal degree
of freedom, the qubit follows an interior gradient-flow path
from the north pole ($z=+1$) to the south pole ($z=-1$),
rather than rotating on the surface as in unitary Rabi oscillations.
Thus the $\phi^4$ instanton of the auxiliary field manifests physically
as a \emph{Bloch-sphere instanton}: a real-time, deterministic tunneling
trajectory connecting the excited and ground states.
This establishes a concrete algebraic and geometric correspondence between
the SU(1,1) structure of the driven qubit and the instanton equations of
a $\phi^4$ field theory.

Most importantly however, we note that under our assumption of the PT-symmetric parametric driving $g(t)=g_0e^{-i\omega_g t}$, the tunneling solution is unidirectional. In this way, our solution represents a truly quantum mechanical form of the unidirectional transport associated with non-Hermitian physics.  In the language of dynamical systems, the ground state becomes a strong attractor within the quantum dynamics of the Bloch ball. While we have considered the particular case of a non-Hermitian parameteric drive which causes the emergence of a strong attractor at the ground state, using the same framework it is easy to generalize this theory to a family of non-Hermitian parametric drives, such that any point on the bloch sphere can be forced to become a strong attractor through the same kind of instanton dynamics. This generalization is given explicitly in the End Matter.

\begin{figure}[t]
\centering
\includegraphics[width=0.5\textwidth]{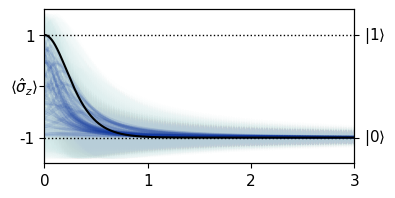}
\caption{The dynamics of the z-component of the spin ($\langle \hat\sigma_z\rangle$) of a qubit under PT-symmetric parametric forcing. Black Line: analytical solution (Eq. \ref{inst}) for initial state $ \hat\sigma_z(0)=1$ and $\langle \hat a \hat a^\dagger\rangle(0)=0$. Dark blue lines show the result of direct numerical simulations of the full quantum Rabi model, starting from one-million random initial conditions, which evenly sample the Bloch sphere, and which randomly samples the initial cavity Boson expectation within $[0,5]$. Light blue shading shows the bounds of quantum uncertainty. For an unknown initial state, the analytical curve represents an outer-bound estimate for the time to reset the unknown state coherently.
}\label{cool}
\end{figure}

So far, we have made mathematical arguments which predict the appearance of an attractor at the ground state of the Bloch sphere representing the qubit spin, given the non-Hermitian (non-unitary) parametric driving of the form $g(t)=g_0e^{-i\omega_g t}$. However, those arguments did involve approximations, so it is important to confirm that this striking phenomenon still appears in the full quantum Rabi model. For such confirmation, we numerically integrate the full quantum Rabi Hamiltonian (Eq. \ref{qrm}) with time dependent complex coupling using the QuTip Python Library. We choose one million randomly distributed points on the Bloch sphere to define a set of initial qubit states, and assuming a random initial cavity photon occupation number (between $[0,5])$, we evolve the Rabi model for each of the one million initial value problems. Fig. \ref{cool} shows the results, with the numerically calculated dynamics of $ \hat\sigma_z$ shown in dark blue lines rendered in low opacity to highlight the overall behavior. That figure also shows the analytical instanton solution given by Eq. \ref{inst} in black, for the qubit starting in the excited state and for the cavity in an initial vacuum state. That figure shows clearly that all initial states rapidly converge to the ground state expectation, with an approximate outer-bound given by the analytical solution for the initial excited state.

In summary, we have introduced a robust, non-Hermitian single-step qubit reset protocol, which exploits the emergence of unidirectional instantons for the coherent one-way tunneling to the ground state. While state of the art reset mechanisms rely on either incoherent jump processes or multistep readout / operation processes, this protocol  does not require coupling to a lossy resonator which can introduce decoherence, and is simultaneously blind to the initial state of the qubit. Further, this protocol relies only on the nature of generic qubits, and on the symmetry of the parametric forcing. While here we have only considered parametric driving which leads to unidirectional tunneling to the ground state, it remains possible that there exist similar drives which cause the emergence of attractors at other points on the Bloch sphere.

Due to the critical importance of qubit cooling and reset techniques to quantum computing and metrology, we expect this work to lead to experimental realizations of non-Hermitian parametric driven systems, and to inspire further theoretical work towards understanding the nature of non-Hermitian driven quantum systems. In particular, we expect to find that unidirectional instanton solutions will be found to be a universal process which allows for general unidirectional coupling processes in fully quantum dynamics.

The author thanks Natalia Berloff and the University of Cambridge Centre for Mathematical Sciences for their generous hospitality while this work began. The author also thanks Eddy Timmermans for his encouragement. This work was supported by Los Alamos National
Laboratory LDRD program grant 20230865PRD3.


\onecolumngrid
\section{End Matter}
\twocolumngrid
\noindent\textit{{Generalization to Arbitrary Attractors -- }}
In the foregoing analysis, the PT-symmetric modulation
$g(t)=g_{0}e^{-i\omega_{g}t}$ generates a non-Hermitian
flow on the Bloch ball whose unique fixed point is the
ground state.
This arises because that drive couples the two counter-rotating
terms of the quantum Rabi Hamiltonian with equal strength
and quadrature phase, enforcing a purely longitudinal
(axial) gradient flow.
More generally, the orientation of this non-Hermitian flow
can be controlled analytically by modifying the complex
trajectory of the drive in the two-dimensional coupling
plane.

We therefore consider a generalized parametric modulation
\begin{equation}
  g(t)=g_{0}\!\left[\cos(\omega_{g}t+\phi_{x})
  + i\,\eta\,\sin(\omega_{g}t+\phi_{y})\right],
  \label{eq:ellipticaldrive}
\end{equation}
which represents an \emph{elliptically polarized} non-Hermitian drive
with eccentricity~$\eta$ and orientation angle
$\Phi=(\phi_{x}-\phi_{y})/2$.
For $\eta=1$ and $\Phi=0$,
Eq.~\eqref{eq:ellipticaldrive} reduces to the circularly
symmetric case analyzed above.
Within the same rotating-wave approximation,
the effective Hamiltonian retains the SU(1,1) structure,
but with a complex coupling vector
$\boldsymbol{\alpha}\!\propto\!
g_{0}(\eta\cos\Phi,\,\eta\sin\Phi,\,1)$
tilted away from the $z$~axis by an angle determined by
$\eta$ and~$\Phi$.
The associated Wei--Norman coefficients satisfy
\begin{equation}
  \dot f_{1}=i\alpha_{z}(1+f_{1}^{2})
  -2i(\alpha_{x}+i\alpha_{y})f_{1},
  \label{eq:generalfn}
\end{equation}
which generalizes the Riccati flow of Eq.~(10) to include
transverse complex coupling components.
Its stationary solutions determine the fixed points of the
non-Hermitian flow,
\begin{equation}
  \langle\boldsymbol{\sigma}\rangle_{\!\infty}
  =-\frac{\boldsymbol{\alpha}}{|\boldsymbol{\alpha}|},
  \label{eq:fixedpoint}
\end{equation}
showing that the attractor lies along the direction of the
complex coupling vector~$\boldsymbol{\alpha}$.

Analytically, increasing the eccentricity $\eta>1$ breaks
the rotational symmetry between the $\sigma_{x}$ and
$\sigma_{y}$ quadratures, thereby fixing a preferred
azimuthal angle~$\phi$ on the Bloch sphere, while rotating
the ellipse by~$\Phi$ tilts the effective gain--loss axis and
sets the attractor’s polar angle~$\theta$.
In this manner, the parameters $(\eta,\Phi)$ define a
continuous two-parameter family of fixed points
$\langle\boldsymbol{\sigma}\rangle_{\!\infty}(\eta,\Phi)$,
forming a smooth manifold of attractors on the Bloch
sphere.
Each corresponds to a distinct instanton trajectory of the
generalized Riccati system~\eqref{eq:generalfn}, and for
each $(\eta,\Phi)$ the qubit undergoes a unidirectional,
monotonic evolution of the same instantonic form as
Eq.~(12), converging toward the corresponding fixed state
defined by Eq.~\eqref{eq:fixedpoint}.

This analysis shows that the Bloch-sphere instanton is not
restricted to ground-state reset/cooling but describes a general
family of non-Hermitian instanton trajectories that
can be continuously steered to any point on the Bloch
sphere, still without knowledge of the initial state, through the parameters $(\eta,\Phi)$.
The PT-symmetric parametric drive therefore defines an
exact, analytical framework for arbitrary coherent state engineering
via deterministic tunneling.

\end{document}